\documentclass[aps,pra,twocolumn,superscriptaddress,nofootinbib]{revtex4-1}
\usepackage{amsmath}
\usepackage{amssymb}
\usepackage{graphicx}
\usepackage{mathrsfs}
\usepackage{ntheorem}
\usepackage{mathtools}
\usepackage{enumerate}
\usepackage{enumitem}
\usepackage{times,txfonts}
\usepackage{subfigure}
\usepackage{upgreek}
\usepackage{tikz}
\usepackage[colorlinks,bookmarksopen,bookmarksnumbered,citecolor=blue, linkcolor=blue, urlcolor=blue]{hyperref}
\newcommand{\ket}[1]{|#1\rangle}
\newcommand{\bra}[1]{\langle #1|}
\newcommand{\Tr}{\mathrm{Tr}}

\newcommand{\abs}[1]{\lvert #1\rvert}

\def\CC{{\rm\kern.24em \vrule width.04em height1.46ex depth-.07ex \kern-.30em C}}
\def\RR{{\rm\kern.24em \vrule width.04em height1.46ex depth-.07ex
\kern-.30em R}}
\def\P{{\rm I\kern-.25em P}}

\begin{document}

\title{Criterion for a state to be distillable via stochastic incoherent operations}

\author{C. L. Liu}
\affiliation{Graduate School of China Academy of Engineering Physics, Beijing 100193, China}
\author{D. L. Zhou}
\email{zhoudl72@iphy.ac.cn}
\affiliation{Institute
  of Physics, Beijing National Laboratory for Condensed Matter
  Physics, Chinese Academy of Sciences, Beijing 100190, China}
\affiliation{School of Physical Sciences, University of Chinese
  Academy of Sciences, Beijing 100049, China}
\affiliation{CAS Central of Excellence in Topological Quantum
  Computation, Beijing 100190, China}
\affiliation{Songshan Lake Materials Laboratory, Dongguan, Guangdong
  523808, China}
\author{C. P. Sun}
\email{suncp@gscaep.ac.cn}
\affiliation{Graduate School of China Academy of Engineering Physics, Beijing 100193, China}
\affiliation{Beijing Computational Science Research Center, Beijing 100193, China}
\date{\today}

\begin{abstract}
Coherence distillation is a basic information-theoretic task in the resource theory of coherence. In this paper, we present the necessary and sufficient conditions under which a mixed state can be distilled into a pure coherent state via stochastic incoherent operations (sIOs). With the help of this result, we further show the following: (i) Any $2$-dimensional coherent state is distillable via sIOs if and only if it is a pure coherent state; (ii) a state $\rho$ is n-distillable via sIOs if and only if it is 1-distillable; and (iii) the set of distillable states via stochastic maximally incoherent operations is identical to the set of distillable states via sIOs. Finally, we analyze the reason why sIO is stronger than stochastic strictly incoherent operations when we use them to distill a coherent state.
\end{abstract}
\maketitle

\section{Introduction}
Quantum coherence is an important feature of quantum mechanics which is responsible for the departure between the classical and quantum world. It is an essential component in quantum information processing \cite{Nielsen}, and plays a central role in various fields, such as quantum computation \cite{Shor,Grover}, quantum cryptography \cite{Bennett}, quantum metrology \cite{Giovannetti}, and quantum biology \cite{Lambert}. Recently, the resource theory of coherence has attracted a growing interest due to the rapid development of quantum information science \cite{Aberg1,Baumgratz,Streltsov}. The resource theory of coherence not only establishes a rigorous framework to quantify coherence but also provides a platform to understand  quantum coherence from a different perspective.

Any quantum resource theory is characterized by two fundamental ingredients, namely, the free states and the free operations \cite{Chitambar}. For the resource theory of coherence, the free states are quantum states which are diagonal in a prefixed reference basis. The free operations are not uniquely specified. Motivated by suitable practical considerations, several free operations were presented \cite{Streltsov}, such as maximally incoherent operations (MIOs) \cite{Aberg1}, incoherent operations (IOs) \cite{Baumgratz}, and strictly incoherent operations (SIOs) \cite{Winter,Yadin}.

In the resource theory of coherence, much effort has been devoted to investigate the coherence distillation \cite{Chitambar}. The coherence distillation is the process that extracts pure coherent states from a mixed state via free operations, and various coherence distillation protocols were proposed.
These protocols can be divided into two different settings: the asymptotic regime
\cite{Yuan,Winter,Lami,Lami1,Zhao} and the one-shot regime \cite{Liu1,Liu2,Liu3,Fang,Regula,Regula1,Chitambar2,Torun,Regula2,Chen,Zhang,Du,Ziwen,Zhu,Du1,Liu4,Chitambar3,Ding}. With the above protocols, the coherence distillation under SIO is well understood. Specifically, in Refs. \cite{Lami,Lami1,Zhao}, the necessary and sufficient conditions for asymptotic distillability are presented. The deterministic coherence distillation is completed in Refs. \cite{Liu1,Du,Zhu,Chitambar2}, and the probabilistic coherence distillation is completed in Refs. \cite{Liu2,Du1,Zhu,Chitambar2,Liu4,Torun}. However, this is not the case for IO. Although the coherence distillation via IO is well understood in the asymptotic case \cite{Winter}, ``less progress has been made for the single copy transformations of mixed coherent states and only isolated results are known" \cite{Streltsov}. A fundamental question of them is, given a mixed coherent state $\rho$, can we transform it into a pure coherent state $\varphi$ via some IO with nonzero probability?

In this paper, we solve the above question. In other words, we find the necessary and sufficient conditions under which a coherent state can be distilled into a pure coherent state via stochastic incoherent operations (sIOs). This result can help us examine some consequences and solve some other open questions: (1) We show that any $2$-dimensional coherent state is distillable via sIO if and only if it is a pure coherent state. (2) We show that a state $\rho$ is n-distillable via sIO if and only if it is 1-distillable, which is an interesting problem in the resource theories. (3) We show that the set of distillable states via stochastic maximally incoherent operations is identical to the set of distillable states via sIO, which is an open question in Refs. \cite{Fang,Fang1}. Finally, we analyze the reason why sIO is stronger than stochastic strictly incoherent operations when we use them to distill a coherent state.

This paper is organized as follows. In Sec.~\ref{II}, we recall some notions of the resource theory of coherence. In Secs.~\ref{III} and ~\ref{IV}, we present the necessary and sufficient conditions under which a mixed state can be distilled into a pure coherent state via sIO and give the explicit distillation protocol. In Sec.~\ref{V}, we present three corollaries about the necessary and sufficient conditions. Section \ref{VI} contains our remarks and conclusions.

\section{Resource theory of coherence}\label{II}

Let $\mathcal {H}$ be the Hilbert space of a $d$-dimensional quantum
system. A particular basis of $\mathcal {H}$ is denoted as
$\{\ket{i}, ~i=1,2,\cdot\cdot\cdot,d\}$, which is chosen according to the physical
problem under consideration. Coherence of a state is then measured
based on the basis. A state is
incoherent if it is diagonal in the basis, i.e., has the form $\delta=\sum_i\delta_i\ket{i}\bra{i}$. Any state which cannot be
written as a diagonal matrix is defined as a coherent state. For the sake of simplicity, we will denote
$\ket{\varphi}\bra{\varphi}$ as $\varphi$ for a
pure state $\ket{\varphi}$, i.e.,
$\varphi:=\ket{\varphi}\bra{\varphi}$ and denote
$\ket{\phi_d}=\frac1{\sqrt{d}}\sum_{i=1}^{d}\ket{i}$ as a
$d$-dimensional maximally coherent state.

For the resource theory of coherence, the most fundamental requirement for the free operations is that: any free operation can only map an incoherent state to an incoherent state. With this requirement and other considerations, several free operations were proposed. We recall three classes of them: the maximally incoherent operations (MIOs) \cite{Aberg1}, the incoherent operations (IOs) \cite{Baumgratz}, and the strictly incoherent operations (SIOs) \cite{Winter,Yadin}, which will be considered in this paper. The MIO is defined to be all operations $\Lambda(\rho)=\sum_{n=1}^N K_n\rho K_n^\dagger$ such that $\Lambda(\delta)\in\mathcal{I}$ for every $\delta\in\mathcal{I}$, where $\mathcal{I}$ is the set of incoherent states. An incoherent operation \cite{Baumgratz} is a completely positive trace-preserving (CPTP) map, expressed as
\begin{eqnarray}
 \Lambda(\rho)=\sum_{n=1}^N K_n\rho K_n^\dagger,
 \end{eqnarray}
with $K_n\mathcal{I}K_n^\dagger\subset \mathcal{I}$ for all $K_n$, i.e., each
$K_n$ transforms an incoherent state into an incoherent state. Hereafter, we will refer to such
a $K_n$ as an incoherent operator, while, a strictly
incoherent operation \cite{Winter, Yadin} is a CPTP map satisfying not only
$K_n\mathcal{I}K_n^\dagger\subset \mathcal{I}$ but also
$K_n^\dagger\mathcal{I}K_n\subset\mathcal{I}$ for all $K_n$, i.e., each
$K_n$ as well $K_n^\dagger$ transforms an incoherent state into an
incoherent state. Hereafter we will refer to such a $K_n$ as a strictly incoherent operator. If we use $\mathcal{S}_\text{MIO}$, $\mathcal{S}_\text{IO}$, and $\mathcal{S}_\text{SIO}$ to represent the sets
of MIO, IO, and SIO, respectively, then they have the
inclusion relationships
\begin{eqnarray}
\mathcal{S}_\text{SIO}\subset\mathcal{S}_\text{IO}\subset\mathcal{S}_\text{MIO}.
\end{eqnarray}

With the notions of MIO, IO, and SIO, we further recall the notion of the stochastic MIO (sMIO) \cite{Fang}, stochastic IO (sIO) \cite{Liu3}, and stochastic SIO (sSIO) \cite{Liu5}. We just recall sIO here, since sMIO and sSIO can be defined similarly. A sIO is constructed by a subset of incoherent operators. Without loss of generality, we denote the subset as $\{K_{1},K_{2},\dots, K_{L}\}$. Otherwise, we may renumber the subscripts of these incoherent operators. Then, a sIO, denoted as $\Lambda_s(\rho)$, is defined by
\begin{equation}
\Lambda_s(\rho)=\frac{\sum_{n=1}^L K_{n}\rho K_{n}^{\dagger}}{\Tr(\sum_{n=1}^LK_{n}\rho K_{n}^{\dagger})},
\label{lams}
\end{equation}
where $\{K_{1},K_{2},\dots, K_{L}\}$ satisfies $\sum_{n=1}^L K_{n}^{\dagger}K_{n}\leq I$. Here, we emphasize that the stochastic transformation with $\sum_{n=1}^L K_{n}^{\dagger}K_{n}\leq I$ means that a copy of $\Lambda_s(\rho)$ may be obtained from a copy of $\rho$ with probability $\text{P}=\Tr(\sum_{n=1}^LK_{n}\rho K_{n}^{\dagger})(\leq1)$. That is, the stochastic transformation runs the risk of failure with certain probability. However, the deterministic transformation with $\sum_{n=1}^N K_{n}^{\dagger}K_{n}= I$ corresponds to the case $\text{P}=1$, i.e., without running any risk of failure. Supposing we use $\mathcal{S}_\text{sMIO}$, $\mathcal{S}_\text{sIO}$, and $\mathcal{S}_\text{sSIO}$ to represent the sets
of sMIO, sIO, and sSIO, respectively, then they will have the
inclusion relationships
\begin{eqnarray}
\mathcal{S}_\text{sSIO}\subset\mathcal{S}_\text{sIO}\subset\mathcal{S}_\text{sMIO}. \label{inclusion}
\end{eqnarray}

\section{Distillation criterion under $\text{sIO}$}\label{III}

To present the main theorem clearly, we first collect the following useful facts and an example:

(i) We say a coherent state $\rho$ is distillable via sIO if we can transform it into some pure coherent state $\varphi$  via some sIO $\Lambda_s$ with nonzero probability. In other words, there is
\begin{eqnarray}
\Lambda_s(\rho)=\ket{\varphi}\bra{\varphi}.
\end{eqnarray}

(ii) We define the coherence support of $\ket{\varphi}$, denoted as $\text{c-supp}(\varphi)$, as the set of incoherent states $\{\ket{i}\}$ which have nonzero overlap with $\ket{\varphi}$, i.e., $\text{c-supp}(\varphi):=\{\ket{i}|~\langle i\ket{\varphi}\neq0\}$. For example, the coherence support of $\ket{\varphi}=\frac1{\sqrt{2}}(\ket{1}+\ket{2})$ is $\{\ket{1},\ket{2}\}$, i.e., $\text{c-supp}(\varphi)=\{\ket{1},\ket{2}\}$.

(iii) Lemma. If we want to see whether a mixed state can be transformed into a pure coherent state via some $ \Lambda_s(\rho)$, we only need to consider the sIO with the form of (see the Appendix for the proof of the Lemma)
\begin{eqnarray}
  \Lambda_s^1(\rho)=\frac{K\rho K^\dag}{\Tr(K\rho K^\dag)}.\label{Step1}
\end{eqnarray}

(iv) Let $\rho$ be a density matrix. Then, we can uniquely write $\rho$ as $\rho=\bigoplus_\mu p_\mu \rho_\mu$ with each $\rho_\mu$ being irreducible \cite{Horn,note}. Here, $\rho_\mu$ is said to be irreducible if we cannot transform it into a block diagonal matrix by using some permutation operation.

With the above facts, let us begin with a simple example to illustrate the general idea of Theorem 1.

(v) Example. Consider a 3-dimensional state
\begin{eqnarray}\label{state}
  \rho=\frac13\begin{pmatrix}
    1&0&1\\
    0&1&0\\
   1&0&1
  \end{pmatrix}.
\end{eqnarray}
We first express $\rho$ as $\rho=\frac23\rho_1\oplus\frac13\rho_2$ with $\rho_1=\frac12(\ket{1}+\ket{3})(\bra{1}+\bra{3})~\text{and}~\rho_2=\ket{2}\bra{2}$, where both $\rho_1$ and $\rho_2$ are irreducible. Second, the spectral decomposition of $\rho_1$ and $\rho_2$ are $\rho_1=\ket{\uplambda}\bra{\uplambda}$ with $\ket{\uplambda}=\frac1{\sqrt{2}}(\ket{1}+\ket{3})$ and $\rho_2=\ket{2}\bra{2}$, respectively. We note that $\rho_1$ is not of full rank and $\rho_2$ is of full rank. Third, since $\rho_1$ is not of full rank, then there is a pure state $\ket{\psi}$, such as $\ket{\psi}=\frac1{\sqrt{2}}(\ket{1}-\ket{3})$, in the null space of $\rho_1$. We note that there is $\text{c-supp}(\psi)\bigcap\text{c-supp}(\uplambda)\neq\emptyset$, where $\text{c-supp}(\psi)=\{\ket{1},\ket{3}\}=\text{c-supp}(\uplambda)$. Then, the sIO can be chosen as $\Lambda_s(\rho)=K\rho K^\dag/\Tr(K\rho K^\dag)$ with
\begin{eqnarray}K=\frac1{\sqrt{2}}(\ket{1}\bra{1}+\ket{2}\bra{3}).
 \end{eqnarray}By direct calculations, we obtain $\Lambda_s(\rho)=\ket{\phi_2}\bra{\phi_2}$ with $\ket{\phi_2}=\frac1{\sqrt{2}}(\ket{1}+\ket{2})$ and $\Tr(K\rho K^\dag)=\frac13$.

With the above discussions, we then present the main result of this paper:

Theorem 1. A coherent state $\rho$ is distillable via sIO if and only if  $\rho$ has the form
\begin{eqnarray}
\rho=\bigoplus_\mu p_\mu \rho_\mu, \label{theorem}
\end{eqnarray}
where each $\rho_\mu$ is irreducible and at least one $ \rho_\mu$ is not of full rank.

Proof.--We first present a criterion for the coherence distillation via sIO.

Suppose there is a sIO $\Lambda(\cdot)$ realizing the transformation $\rho$ into a pure coherent state $\varphi$. Then, we have
 \begin{eqnarray}
 K\rho K^\dag=p\ket{\varphi}\bra{\varphi},\label{sio1}
 \end{eqnarray}
where $p=\Tr( K\rho K^\dag)$ is nonzero.
Let $\{p_\mu,\varphi_\mu\}$ be an arbitrary pure-state ensemble decomposition of $\rho$, i.e., $\rho=\sum_\mu p_\mu\ket{\varphi_\mu}\bra{\varphi_\mu}$.
 We then obtain
 \begin{eqnarray}
 K\left(\sum_\mu p_\mu\ket{\varphi_\mu}\bra{\varphi_\mu}\right) K^\dag=p\ket{\varphi}\bra{\varphi}. \label{sio2}
 \end{eqnarray}
From Eq. (\ref{sio2}) and $p=\Tr( K\rho K^\dag)$ being nonzero, there is $\ket{\varphi}\bra{\varphi}=\sum_\mu \frac{p_\mu}pK\ket{\varphi_\mu}\bra{\varphi_\mu}K^\dag$. For each $K\ket{\varphi_\mu}\bra{\varphi_\mu}K^\dag$, there are two cases to consider. The first case is  $K\ket{\varphi_\mu}\bra{\varphi_\mu}K^\dag=\textbf{0}$ and the second is $K\ket{\varphi_\mu}\bra{\varphi_\mu}K^\dag\neq\textbf{0}$. For the latter case,
since pure states are extreme points of the set of states \cite{note1}, there are $K\ket{\varphi_\mu}\bra{\varphi_\mu}K^\dag=q_\mu\ket{\varphi}\bra{\varphi}$, where $q_\mu=\Tr(K\ket{\varphi_\mu}\bra{\varphi_\mu}K^\dag)$. Here, we should note that, since $p\neq0$ in Eq. (\ref{sio2}), there is $K\ket{\varphi_\mu}\bra{\varphi_\mu}K^\dag=q_\mu\ket{\varphi}\bra{\varphi}$ with $q_\mu\neq0$ for some $\mu$, i.e., $q_\mu$ cannot be zero at the same time. Thus, there are
 \begin{eqnarray}\label{ns}
 K\ket{\varphi_\mu}\bra{\varphi_\mu}K^\dag=q_\mu\ket{\varphi}\bra{\varphi}~\text{or}~
 K\ket{\varphi_\mu}\bra{\varphi_\mu}K^\dag=\bf{0}, \label{nesu}
 \end{eqnarray}
 where $q_\mu=\Tr\left(K\ket{\varphi_\mu}\bra{\varphi_\mu}K^\dag\right)\neq0$ and $\bf{0}$ is the null matrix for all $\mu$. The relations in Eq. (\ref{nesu}) provide a criterion for the distillability of $\rho$.

With the above criterion, we next prove the \emph{if} part of the theorem, i.e., if the state $\rho$ satisfying the conditions in the theorem, then it is distillable via sIO.

To see this, let us first recall the structure of incoherent operators. For an incoherent operator, there is at most one nonzero element in each column of $K$ \cite{Yao, Du, Winter}. Thus, it is direct to obtain that if $K$ is an incoherent operator, then there is
\begin{eqnarray}
K=\sum_i\ket{i}\bra{\phi_i},
\end{eqnarray}
where $\{\ket{\phi_i}\}$  has the disjoint coherence support. Let $\rho=\bigoplus_\mu p_\mu \rho_\mu$ with each $\rho_\mu$ being irreducible and suppose $\rho_\mu$, one of $\{\rho_\mu\}$, is not of full rank. Let $\rho_\mu=\sum_j\uplambda_j^\mu\ket{\uplambda_j^\mu}\bra{\uplambda_j^\mu}$ be the spectral decomposition of $\rho_\mu$ and let $\bigcup_{j}\text{c-supp}(\uplambda_j^\mu)=\{\ket{i^\mu_1},\cdots,\ket{i^\mu_{n_\mu}}\}$, where $n_\mu$ is the number of the elements of the set $\{\ket{i^\mu_1},\cdots,\ket{i^\mu_{n_\mu}}\}$. Since $\rho_\mu$ is not of full rank, there is some $\ket{\psi}\in\text{span}\{\ket{i^\mu_1},\cdots,\ket{i^\mu_{n_\mu}}\}$ in the null space of $\rho_\mu$ \cite{note2}. In other words, let $\ket{\psi}=c_1\ket{i^\mu_1}+\sum_{l=2}^{n_\mu}c_l\ket{i^\mu_l}$. Then there are
\begin{eqnarray}
\text{c-supp}(\psi)\bigcap\left(\cup_{j}\text{c-supp}(\uplambda_j^\mu)\right)\neq\emptyset. \label{union}
\end{eqnarray}
and
\begin{eqnarray}
\langle\uplambda_j^\mu\ket{\psi}=0 \label{orth}
\end{eqnarray}
for all $j$. Furthermore, we should note that since $\rho_\mu$ is an irreducible matrix,  $\ket{\psi}$ cannot be an incoherent state. To see this, let $\ket{\psi}=\ket{i^\mu_l}$ be an incoherent state. Then, from the relations in Eq. (\ref{orth}), we obtain that $\text{c-supp}(i^\mu_l)\bigcap\left(\bigcup_{j}\text{c-supp}(\uplambda_j^\mu)\right)=\emptyset$. This implies that $\rho_\mu$ is not irreducible and we arrive at a contradiction.

Without loss of generality, suppose that there is $c_1\neq0$. Let $\ket{\uplambda_j^\mu}=\sum_{s=1}^{n_\mu}c_s^j\ket{i_s^\mu}$. Then, $\langle\uplambda_j^\mu\ket{\psi}=0$ for all $j$ implies that $\sum_{l=1}^{n_\mu}{c_l^j}^*c_l=0$ for all $j$, i.e., there are ${c_1^j}^*c_1=-\sum_{l=2}^{n_\mu}{c_l^j}^*c_l$ for all $j$. Let us consider the incoherent operator
\begin{eqnarray}
K=c_1\ket{1}\bra{i^\mu_1}-\ket{2}\bra{\psi_1},\label{Kraus}
\end{eqnarray}
where $\ket{\psi_1}:=\sum_{l=2}^{n_\mu}c_l\ket{i^\mu_l}$.
From Eqs. (\ref{orth}) and (\ref{Kraus}), we immediately obtain, for all $j$, $K\ket{\uplambda_j^\mu}={c_1^j}^*c_1\ket{1}-\sum_{l=2}^{n_\mu}{c_l^j}^*c_l\ket{2}
={c_1^j}^*c_1\left(\ket{1}+\ket{2}\right)$, i.e., there are
\begin{eqnarray}
K\ket{\uplambda_j^\mu}\bra{\uplambda_j^\mu}K^\dag=p_j^\mu\ket{\phi_2}\bra{\phi_2}, \label{trans}
\end{eqnarray}
where $p_j^\mu=\Tr\left(K\ket{\uplambda_j^\mu}\bra{\uplambda_j^\mu}K^\dag\right)$ and $\ket{\phi_2}=\frac1{\sqrt{2}}(\ket{1}+\ket{2})$.
Since $\rho_\mu$ is irreducible and $\bigcup_{j}\text{c-supp}(\uplambda_j^\mu)=\{\ket{i^\mu_1},\cdots,\ket{i^\mu_{n_\mu}}\}$, we could obtain that $p_j^\mu\neq0$ for some $j$. To see this, let $p_j^\mu=0$ for every $j$. Then, from Eq. (\ref{trans}), there are $K\ket{\uplambda_j^\mu}\bra{\uplambda_j^\mu}K^\dag=\textbf{0}$. This further implies that, for every $j$, there are ${c_1^j}^*c_1=0$. Since $c_1\neq0$, there are ${c_1^j}^*=0$, i.e., $c_1^j=0$ for all $j$. This means that $\bra{i^\mu_1}\rho_\mu\ket{i^\mu_1}=0$. Since $\rho_\mu$ is a positive semi-definite matrix, then $\bra{i^\mu_l}\rho_\mu\ket{i^\mu_1}=0=\bra{i^\mu_1}\rho_\mu\ket{i^\mu_l}$ for all $l=1,\cdots,n_\mu$. However, this is impossible since $\rho_\mu$ is irreducible.

The above discussions imply that, for a state $\rho=\bigoplus_\mu p_\mu \rho_\mu$, if there is a $\rho_\mu$ satisfying the conditions in the theorem, then we can always transform it into a 2-dimensional maximally coherent state via the $K$ in Eq. (\ref{Kraus}) with nonzero probability. This completes the \emph{if} part of the theorem.

Finally, we prove the \emph{only if} part of the theorem, i.e., for $\rho=\bigoplus_\mu p_\mu \rho_\mu$, if each irreducible $\rho_\mu$ is of full rank, then $\rho$ is not distillable via any sIO.

To this end, we will show that if we can transform such a state into a pure coherent one via some sIO, then we can transform an incoherent state into a pure coherent state. To see this, from Eq. (\ref{ns}), if we can transform $\rho$, with a pure-state ensemble decomposition $\rho=\sum_jp_j\ket{\varphi_j}\bra{\varphi_j}$, into $\ket{\varphi}$ via some sIO, we then obtain
 \begin{eqnarray}\label{ns_1}
 K\ket{\varphi_j}\bra{\varphi_j} K^\dag=q_j\ket{\varphi}\bra{\varphi}, \label{eq3}
 \end{eqnarray}
 where $q_j=\Tr\left(K\ket{\varphi_j}\bra{\varphi_j}K^\dag\right)$ with $\sum_jq_j\neq0$. If there is another $\rho^\prime=\sum_ip_i^\prime\ket{\psi_i}\bra{\psi_i}$ with the same range as $\rho$, then there are $\ket{\psi_i}=\sum_jc_{ij}\ket{\varphi_j}$ for all $i$. Here, the range of a matrix $A$, denoted as $\mathbb{R}(A)$, is $\mathbb{R}(A)=\{\ket{y}:\ket{y}=A\ket{x}\}$ \cite{Horn}.
Thus, by using the relations in Eq. (\ref{eq3}), we derive
 \begin{eqnarray}
 K\ket{\psi_i}&&=K\sum_jc_{ij}\ket{\varphi_j}=\sum_jc_{ij}K\ket{\varphi_j}\nonumber\\&&=\sum_jc_{ij}\sqrt{q_j}e^{i\theta_j}\ket{\varphi}=c_i^\prime\ket{\varphi},
 \end{eqnarray}
 where $c_i^\prime:=\sum_jc_{ij}\sqrt{q_j}e^{i\theta_j}$ and $\theta_j$ are some global phase factors from Eq. (\ref{eq3}).
 Since $\mathbb{R}(\rho)$ is identical with $\mathbb{R}(\rho^\prime)$, we also obtain that there are $\ket{\varphi_j}=\sum_id_{ji}\ket{\psi_i}$. These relations imply that at least one $c_i^\prime$ is nonzero, since otherwise all the $q_j$ in Eq. (\ref{eq3}) are zero.

Suppose $\rho_\mu=\sum_{j=1}^{n_\mu}\uplambda_j^\mu\ket{\uplambda_j^\mu}\bra{\uplambda_j^\mu}$ is the spectral decomposition of $\rho_\mu$. Since $\rho_\mu$ is irreducible and is of full rank, then the state
\begin{eqnarray}
\rho_\mu^\prime=\frac1{n_\mu}\sum_{j=1}^{n_\mu}\ket{\uplambda_j^\mu}\bra{\uplambda_j^\mu}
\end{eqnarray}
 is an incoherent state and has the same range as $\rho_\mu$. Thus, if the state $\rho_\mu$ is distillable via some sIO, then, by the above discussions, this means that $\rho^\prime_\mu$ is also distillable via the same sIO. This is impossible since an incoherent operator cannot transform an incoherent state into any coherent state. Hence, if an irreducible $\rho_\mu$ is of full rank, then it is not distillable.  This further implies that given $\rho=\bigoplus_\mu p_\mu \rho_\mu$ with each $\rho_\mu$ being irreducible and being of full rank,  then it is not distillable. This completes the \emph{only if} part of the theorem. ~~~~~~~~~~ ~~~~~~~~~~ ~~~~~~~~~~$\blacksquare$

 \section{Elementary steps and examples} \label{IV}

With Theorem 1, one can directly examine whether a given mixed state is distillable or not via sIO. To this end, we need to carry out the following three steps:

(i) Transform $\rho$ into the form $\rho=\bigoplus_\mu p_\mu \rho_\mu$ with each $\rho_\mu$ being irreducible.

To this end, we could use the connection between a matrix and its directed graph in Ref. \cite{Horn} and  the numerical algorithm presented in Ref. \cite{Nuutila}. By this algorithm, we could write $\rho$ as $\bigoplus_\mu p_\mu \rho_\mu$ with each $\rho_\mu$ being irreducible.

(ii) Examine whether each $\rho_\mu$ is of full rank or not.

To this end, we just express each $\rho_\mu$ in terms of spectral decomposition. If there is some $\rho_\mu$ that is not of full rank, then $\rho$ is distillable. If all $\rho_\mu$ are of full rank, then $\rho$ is not distillable under sIO.

(iii) Construct the sIO.

To this end, suppose $\rho_\mu$ is not of full rank. Then, there is a pure coherent state $\psi$ in the null space of $\rho_\mu$. Without loss of generality, let $\ket{\psi}=c_1\ket{i^\mu_1}+\sum_{l=2}^{n_\mu}c_l\ket{i^\mu_l}$ which satisfies Eqs. (\ref{union}) and (\ref{orth}). Then the sIO can be chosen as $\Lambda_s(\rho)=K\rho K^\dag/\Tr(K\rho K^\dag)$ with $K=c_1\ket{1}\bra{i^\mu_1}-\ket{2}\sum_{l=2}^{n_\mu}c_l\bra{i^\mu_l}$.

In the following, we give an example to illustrate how to use Theorem 1 and the above steps.

Example. Let us consider the 4-dimensional state
\begin{eqnarray}\label{state}
  \rho=\begin{pmatrix}
    \frac14&0&\frac1{2\sqrt{5}}&\frac1{4\sqrt{5}}\\
    0&\frac14&-\frac1{4\sqrt{5}}&\frac1{2\sqrt{5}}\\
   \frac1{2\sqrt{5}}&-\frac1{4\sqrt{5}}&\frac14&0\\
    \frac1{4\sqrt{5}}&\frac1{2\sqrt{5}}& 0&\frac14
  \end{pmatrix}.
\end{eqnarray}
By step (i), it is straightforward to examine that $\rho$ is irreducible. By step (ii), $\rho=\frac12\left(\ket{\uplambda_1}\bra{\uplambda_1}+\ket{\uplambda_2}\bra{\uplambda_2}\right)$ is the spectral decomposition of $\rho$, where
\begin{eqnarray}
\ket{\uplambda_1}&&=\frac1{5\sqrt{2}}(4,3,\sqrt{5},2\sqrt{5})^t,\\
\ket{\uplambda_2}&&=\frac1{5\sqrt{2}}(-3,4,-2\sqrt{5},\sqrt{5})^t.
\end{eqnarray}
This implies that the rank of $\rho$ is 2 ($<4$). Thus, the state $\rho$ is irreducible and is not of full rank. By step (iii), we can find a vector
\begin{eqnarray}
\ket{\psi}=\left(\sqrt{0.5},0,-\sqrt{0.4},-\sqrt{0.1}\right)^T
\end{eqnarray}
satisfying $\text{c-supp}(\psi)\bigcap\left(\cup_{j=1}^2\text{c-supp}(\uplambda_j)\right)\neq\emptyset$ and $\ket{\psi}\in\text{null}(\rho)$, where $\text{c-supp}(\psi)=\{\ket{1},\ket{3},\ket{4}\}$ and $\text{c-supp}(\uplambda_1)=\text{c-supp}(\uplambda_2)=\{\ket{1},\ket{2},\ket{3},\ket{4}\}$. Then, by Eq. (\ref{Kraus}), the corresponding incoherent operator is
\begin{eqnarray}
  K=\sqrt{0.5}\ket{1}\bra{1}+\sqrt{0.4}\ket{2}\bra{3}+\sqrt{0.1}\ket{2}\bra{4}.
\end{eqnarray}
By direct calculations, we obtain $\Lambda_s(\rho)=\ket{\phi_2}\bra{\phi_2}$ and $\Tr(K\rho K^\dag)=\frac14$.

\section{Some consequences of Theorem 1}\label{V}

Theorem 1 provides the necessary and sufficient conditions for a state to be distillable via sIO. In the following, we present several corollaries of the Theorem 1.

(1) It is direct to obtain that a $2$-dimensional coherent state is distillable via sIO if and only if it is a pure coherent state. To see this, let $\rho$ be a 2-dimensional state. Then, the rank of $\rho$ is 1 or 2. If the rank of $\rho$ is 1, then it is a pure coherent state and it is distillable. If the rank of $\rho$ is 2, then it is a state of full rank. By using Theorem 1, it cannot be distillable via any sIO.

(2) The second one is the relation between n-distillability and 1-distillability of a coherent state via sIO. Here, we say that a state $\rho$ is n-distillable if we can transform $n$ copies of $\rho$, $\rho^{\otimes n}$, into a pure coherent state with nonzero probability. We should note that, for a general resource theory, the n-distillability of a state is not necessary equivalent to the 1-distillability, such as entanglement \cite{Watrous}. However, for the resource theory of coherence, n-distillability is equivalent to the 1-distillability both in the asymptotic case and in probabilistic case. On the one hand, as shown in Ref. \cite{Winter}, in the asymptotic case, the distillable coherence of $\rho$ via IO is
\begin{eqnarray}
C_d(\rho)=S(\Delta\rho)-S(\rho) \label{entropy}
\end{eqnarray}
where $S(\rho):=-\Tr(\rho\ln\rho)$ is the von Neumann entropy, the logarithm is to base 2, and $\Delta(\cdot)=\sum_i\ket{i}\bra{i}(\cdot)\ket{i}\bra{i}$ is the completely dephasing channel. Since $C_d(\rho)$ is additivity, i.e., $C_d(\rho_1\otimes\rho_2)=C_d(\rho_1)+C_d(\rho_2)$, we immediately obtain that n-distillability is identical to 1-distillability in the asymptotic case. On the other hand, from Theorem 1, we obtain that n-distillability and 1-distillability of a coherent state via sIO. To see this, it is direct to examine that if $\rho_\mu$ is of full rank if and only if $\rho^{\otimes n}$ is also of full rank. By Theorem 1, we then have the following:

Corollary 1.  A state $\rho$ is n-distillable via sIO if and only if it is 1-distillable.

(3) The third one is the set of distillable states via sMIO is identical to the set of distillable states via sIO. The necessary and sufficient condition for a state to be distilled via sMIO has attracted much attention \cite{Fang,Fang1}. Although a necessary condition for a state to be distillable via sMIO is provided \cite{Fang,Fang1} which says that it is impossible to distill a full rank coherent state via sMIO, the necessary and sufficient condition for a state to be distilled via sMIO is still an open question. With Theorem 1, we can provide an answer to this open question. This arrives at the following corollary:

Corollary 2. A coherent state $\rho$ is distillable via sMIO if and only if the state $\rho$ has the form $\rho=\bigoplus_\mu p_\mu \rho_\mu$, where each $\rho_\mu$ is irreducible and at least one $ \rho_\mu$ is not of full rank.

We now prove the Corollary 2. Since there is $\mathcal{S}_\text{sIO}\subset\mathcal{S}_\text{sMIO}$, we obtain that any state can be distilled via sIO can also be distilled via sMIO. This implies if $\rho$ has the form $\rho=\bigoplus_\mu p_\mu \rho_\mu$, where each $\rho_\mu$ is irreducible and at least one $ \rho_\mu$ is not of full rank, then it is distillable under sMIO. On the other hand, since it is impossible to distill a full-rank coherent state via sMIO, this completes the proof of the \emph{only if} part. ~~~~~~~~~~ ~~~~~~~~~~$\blacksquare$

\section{Remarks and Conclusions}\label{VI}

Before concluding, we analyze the reason why sIO is stronger than stochastic strictly incoherent operations when we use them to distill a coherent state.

The operational difference between $\text{sIO}$ and $\text{sSIO}$ has attracted much attention in Refs. \cite{Chitambar2,Liu3,Liu4}.
From Theorem 1 and a result in Ref. \cite{Liu2} which says that a coherent state $\rho$ is distillable via sSIO if and only if it contains a rank-one sub-matrix with its dimension greater than or equal to 2, we obtain that the coherence distillation via sIO is in sharp contrast with the coherence distillation via sSIO. Here, we provide a reason why sIOs are generally stronger than sSIOs in coherence distillation.

To this end, we first recall the mathematical structure of them: an incoherent operator has the form
\begin{eqnarray}
K=\sum_i\ket{i}\bra{\psi_i},\label{inop}
\end{eqnarray}
where $\{\ket{\psi_i}\}$  have the disjoint coherence support, while a strictly incoherent operator has the form \cite{Winter}
\begin{eqnarray}
K_s=\sum_i\ket{i}\bra{\pi(i)},\label{sinop}
\end{eqnarray}
where $\pi$ is a permutation. To compare the difference between $\text{sIO}$ and $\text{sSIO}$, we further assume that $\ket{\psi_i}$ in Eq. (\ref{inop}) are not incoherent states at the same time. Suppose we obtain a pure coherent state from $\rho=\sum_\mu p_\mu\ket{\varphi_\mu}\bra{\varphi_\mu}$ via $K$ and $K_s$, respectively.  Then, from the analysis around Eq. (\ref{nesu}), we obtain
 \begin{eqnarray}\label{ns1}
 K\ket{\varphi_\mu}\bra{\varphi_\mu}K^\dag=p_\mu\ket{\varphi}\bra{\varphi}~\text{or}~
 K\ket{\varphi_\mu}\bra{\varphi_\mu}K^\dag=\bf{0}, \label{nesu_i}
 \end{eqnarray}
 and similarly, we can also get that
\begin{eqnarray}\label{ns2}
 K_s\ket{\varphi_\mu}\bra{\varphi_\mu}K_s^\dag=p_\mu\ket{\varphi}\bra{\varphi}~\text{or}~
 K_s\ket{\varphi_\mu}\bra{\varphi_\mu}K_s^\dag=\bf{0}. \label{nesu_s}
 \end{eqnarray}
From the condition $ K\ket{\varphi_\mu}\bra{\varphi_\mu}K^\dag=\bf{0}$ and the form of $K$ Eq. (\ref{inop}), we then obtain $\bra{\psi_i}\varphi_\mu\rangle=0$ for all $i$. Since $\{\ket{\psi_i}\}$  have the disjoint coherence support, it is direct to obtain that there is at least one pure coherent state $\ket{\psi}\in\text{null}(K)$ unless $K$ is a strictly incoherent operator. This implies that if there is a $\ket{\varphi_\mu}$ such that $K\ket{\varphi_\mu}\bra{\varphi_\mu}K^\dag=p_\mu\ket{\varphi}\bra{\varphi}$, then any state that has the same range as
\begin{eqnarray}
\text{span}\{\ket{\varphi_\mu},\ket{\psi}\}
\end{eqnarray}
can be distilled via the sIO $\Lambda_s(\rho)=K\rho K^\dag/\Tr(K\rho K^\dag)$. However, for a strictly incoherent operator, this is not the case. To see this, suppose $ K_s\ket{\varphi_\mu}\bra{\varphi_\mu}K_s^\dag=\bf{0}$. Then, as $K_s$ has the form in Eq. (\ref{nesu_s}), we immediately derive
\begin{eqnarray}
\text{c-supp}(\varphi_\mu)\bigcap \{\pi(i)\}=\emptyset.
\end{eqnarray}
Thus the condition $ K_s\ket{\varphi_\mu}\bra{\varphi_\mu}K_s^\dag=p_\mu\ket{\varphi}\bra{\varphi}$ implies the rank of $\text{P}_\pi\rho\text{P}_\pi$ is one, where $\text{P}_\pi=\sum_i\ket{\pi(i)}\bra{\pi(i)}$. From the proof of Theorem 1 and the above discussions, it is the difference between the null space of the incoherent operator and the null space of the strictly incoherent operator that leads to their different ability in coherence distillation.

The difference between the null space of incoherent operators and strictly incoherent operators also demonstrates the difference between IO and SIO in distinguishing a set of orthogonal states $\{\ket{\psi_i}\}$. (See also Ref. \cite{Yadin}, where the authors discussed the case when $\{\ket{\psi_i}\}$ is a basis.) Here, we say that we can distinguish a set of orthogonal states $\{\ket{\psi_i}\}$ via a set of measurement operators $\{M_n\}$ with $\sum_nM_n=\mathbb{I}$ if, for any $\ket{\psi_i}$, there is an $M_n$ such that
\begin{eqnarray}
\bra{\psi_i}M_n\ket{\psi_i}=1~\text{and}~\bra{\psi_j}M_n\ket{\psi_j}=0~\text{with}~ j\neq i. \label{disting}
\end{eqnarray}
Thus, if there is $\text{c-supp}(\psi_i)\bigcap\text{c-supp}(\psi_j)\neq\emptyset$ for $i\neq j$, then we cannot distinguish them via SIO.  However, this task can be achieved by using IO. We may just construct the following measurement operators $M_n=K_n^\dag K_n$ with
\begin{eqnarray}
K_n=\ket{n}\bra{\psi_n}.
\end{eqnarray}
It is straightforward to examine that $M_n$ can distinguish the set of states $\{\ket{\psi_i}\}$ by using Eq. (\ref{disting}). From the above discussions, we infer the following corollary.

Corollary 3.  For a set of orthogonal states $\{\ket{\psi_i}\}$, we can always distinguish them via IO. However, we can distinguish them via SIO if and only if there is
$\text{c-supp}(\psi_i)\bigcap\text{c-supp}(\psi_j)\neq\emptyset$ for all $i\neq j$.

To summarize, we have presented the criterion for a state to be distillable under sIO.
The main finding is presented as Theorem 1. Theorem 1 gives the necessary and sufficient conditions under which a mixed state can be distilled into a pure coherent state via sIO. With the help of this result, we further show that a state $\rho$ is n-distillable via sIO if and only if it is 1-distillable in Corollary 1, and the set of distillable states via sMIO is identical to the set of distillable states via sIO in Corollary 2. Finally, we analyze the reason why sIO is stronger than sSIO when we use them to distill a pure state. 

\section*{Acknowledgments}

This work is supported by the National Natural Science Foundation of China (NSFC) (Grants No. 12088101, No. 11775300, No. 12075310, No. U1930403, and No. U1930402). C.L.L acknowledges support from the China Postdoctoral Science Foundation Grant No. 2021M690324.

\section*{Appendix}\label{Appendix}

In this appendix, we show that if we want to see whether a mixed state can be transformed into a pure coherent state via some $ \Lambda_s(\rho)$, we only need to consider the sIO with the form of
$\Lambda_s^1(\rho)=K\rho K^\dag/\Tr(K\rho K^\dag).$

To see this, suppose we could transform a mixed state $\rho$ into a pure coherent state $\varphi$ via some sIO $\Lambda_s$. Then there is
$\Lambda_s(\rho)=\sum_{n=1}^L K_{n}\rho K_{n}^{\dagger}/\Tr(\sum_{n=1}^LK_{n}\rho K_{n}^{\dagger})=\varphi.$
Since pure states are extreme points of the set of states \cite{note1}, there are
$K_{n}\rho K_{n}^{\dagger}=p_n\varphi$
for all $n=1,...,L$, where $p_n=\Tr(K_{n}\rho K_{n}^{\dagger})$. Furthermore, since we must transform $\rho$ into $\varphi$ via $\Lambda_s$ with nonzero probability, there is at least one $p_n\neq0$. This means that there must be $K_n\rho K_n^\dag/\Tr(K_n\rho K_n^\dag)= \varphi$
for some $n$. On the other hand, if we have realized the transformation
$\Lambda_s^1(\rho)=K_n\rho K_n^\dag/\Tr(K_n\rho K_n^\dag)=\varphi,$
then there is a sIO realizing the transformation $\Lambda_s(\rho)=\varphi$ since $\Lambda_s^1(\cdot)$ is a sIO.


\begin{thebibliography}{99}
\bibitem{Nielsen} M. A. Nielsen and I. L. Chuang, \emph{Quantum Computation and Quantum Information}, (Canbrudge University Press, Cambridge, U.K., 2000).
\bibitem{Shor} P. W. Shor, SIAM J. Comput. \textbf{26}, 1484 (1997).
\bibitem{Grover} L. K. Grover, Phys. Rev. Lett. \textbf{79}, 325 (1997).
\bibitem{Bennett} C. H. Bennett and G. Brassard, Theor. Comput. Sci. \textbf{560}, 7 (2014).
\bibitem{Giovannetti} V. Giovannetti, S. Lloyd, and L. Maccone,  Science \textbf{306}, 1330 (2004).
\bibitem{Lambert} N. Lambert, Y.-N. Chen, Y.-C. Cheng, C.-M. Li,  G.-Y. Chen, and F. Nori, Nat. Phys.  \textbf{9}, 10 (2013).
\bibitem{Aberg1} J. {\AA}berg, arXiv:quant-ph/0612146.
\bibitem{Baumgratz} T. Baumgratz, M. Cramer, and M. B. Plenio, Phys. Rev. Lett. \textbf{113}, 140401 (2014).
\bibitem{Streltsov} A. Streltsov, G. Adesso, and M. B. Plenio, Rev. Mod. Phys. \textbf{89}, 041003 (2017).
\bibitem{Chitambar} E. Chitambar and G. Gour, Rev. Mod. Phys. \textbf{91}, 025001 (2019).
\bibitem{Winter} A. Winter and D. Yang, Phys. Rev. Lett. \textbf{116}, 120404 (2016).
\bibitem{Yadin} B. Yadin, J. Ma, D. Girolami, M. Gu, and V. Vedral, Phys. Rev. X \textbf{6}, 041028 (2016).
\bibitem{Yuan} X. Yuan, H. Zhou, Z. Cao, and X. Ma, Phys. Rev. A \textbf{92}, 022124 (2015).
\bibitem{Lami1} L. Lami, B. Regula, and G. Adesso, Phys. Rev. Lett. \textbf{122}, 150402 (2019).
\bibitem{Zhao} Q. Zhao, Y. Liu, X. Yuan, E. Chitambar, and A. Winter, IEEE Trans. Inf. Theory \textbf{65}, 6441 (2019).
\bibitem{Lami} L. Lami, IEEE Trans. Inf. Theory \textbf{66}, 2165 (2020).
\bibitem{Chitambar2} E. Chitambar and G. Gour, Phys. Rev. A \textbf{94}, 052336 (2016).
\bibitem{Liu1} C. L. Liu and D. L. Zhou, Phys. Rev. Lett. \textbf{123}, 070402 (2019).
\bibitem{Du} S. Du, Z. Bai, and Y. Guo, Phys. Rev. A \textbf{91}, 052120 (2015).
\bibitem{Zhu} H. Zhu, Z. Ma, Z. Cao, S. M. Fei, and V. Vedral, Phys. Rev. A \textbf{96} 032316 (2017).
\bibitem{Du1} S. Du, Z. Bai, and X. Qi, Quantum Inf. Comput. \textbf{15}, 1307 (2015).
\bibitem{Liu2} C. L. Liu and D. L. Zhou, Phys. Rev. A \textbf{101}, 012313 (2020).
\bibitem{Torun} G. Torun, L. Lami, G. Adesso, and A. Yildiz, Phys. Rev. A \textbf{99}, 012321 (2019).
\bibitem{Liu4} C. L. Liu and C. P. Sun,  Phys. Rev. Research \textbf{3}, 043220 (2021).
\bibitem{Regula} B. Regula, L. Lami, and A. Streltsov, Phys. Rev. A \textbf{98}, 052329 (2018).
\bibitem{Regula1} B. Regula, K. Fang, X. Wang, and G. Adesso, Phys. Rev. Lett. \textbf{121}, 010401 (2018).
\bibitem{Fang} K. Fang, X. Wang, L. Lami, B. Regula, and G. Adesso, Phys. Rev. Lett. \textbf{121}, 070404 (2018).
\bibitem{Chitambar3} E. Chitambar, Phys. Rev. A  \textbf{97}, 050301(R) (2018).
\bibitem{Ziwen}Z.-W. Liu, K. F. Bu, and R. Takagi, Phys. Rev. Lett. \textbf{123}, 020401 (2019).
\bibitem{Chen} S. Chen, X. Zhang, Y. Zhou, and Q. Zhao, Phys. Rev. A \textbf{100}, 042323 (2019)
\bibitem{Regula2}B. Regula, V. Narasimhachar, F. Buscemi, and M. Gu,  Phys. Rev. Research \textbf{2}, 013109 (2020).
\bibitem{Zhang}S. Zhang, Y. Luo, L.-H. Shao, Z. Xi, and H. Fan, Phys. Rev. A \textbf{102}, 052405 (2020).
\bibitem{Liu3} C. L. Liu and D. L. Zhou, Phys. Rev. A \textbf{102}, 062427 (2020).
\bibitem{Ding} Q. Ding and Q. Liu, J. Phys. A \textbf{55} 105301 (2022).
\bibitem{Fang1} K. Fang and Z.-W. Liu, Phys. Rev. Lett. \textbf{125}, 060405 (2020).
\bibitem{Liu5} C. L. Liu, Y.-Q. Guo, and D. M. Tong, Phys. Rev. A \textbf{96}, 062325 (2017).
\bibitem{note} To show this, we only need to use two results in Ref. \cite{Horn}: (i) For a matrix M, there is a unique directed graph \cite{Horn} corresponding to it; and (ii) a matrix M is irreducible if and only if its directed graph G is strongly connected.
\bibitem{Horn} R. A. Horn and C. R. Johnson, \emph{Matrix Analysis}, (Cambridge University Press, Cambridge, U.K. 1990).
\bibitem{note1} An extreme point of a convex set, $S$, is a point $\textbf{x}\in S$, with the property
that if $\textbf{x}=p\textbf{x}_1+(1-p)\textbf{x}_2$ with $\textbf{x}_1,\textbf{x}_2\in S$ and $0<p<1$
then there is $\textbf{x}=\textbf{x}_1=\textbf{x}_2$. Here, we could prove pure states are extreme points of the set of states. To this end, if $\sum_np_n\ket{\varphi_n}\bra{\varphi_n}=\ket{\varphi}\bra{\varphi}$ with $0<p_n<1$ and $\sum_np_n=1$, then there is $\sum_np_n\abs{\langle\varphi_n\ket{\varphi}}^2=1$. This implies that $\abs{\langle\varphi_n\ket{\varphi}}^2=1$, i.e., $\ket{\varphi_n}\bra{\varphi_n}=\ket{\varphi}\bra{\varphi}$, for all $n$. This means pure states are extreme points of the set of states.
\bibitem{Yao} Y. Yao, X. Xiao, L. Ge, and C. P. Sun, Phys. Rev. A \textbf{92}, 022112 (2015).
\bibitem{note2} Let $\mathcal{S}$ be a subset of a vector space. If $\mathcal{S}\neq\emptyset$, then $\text{span}\mathcal{S}$ is the set of linear combinations of elements of $\mathcal{S}$. A fundamental principle in linear algebra is that a linearly independent list of vectors in a finite-dimensional vector space can be extended to a basis \cite{Horn}. Since $\rho_\mu$ is not of full rank, the set $\{\ket{\uplambda_j^{\mu}}\}_j$ is orthogonal but cannot comprise a basis for $\mathcal{H}_\mu$ spanned by $\{\ket{i^\mu_1},\cdots,\ket{i^\mu_{n_\mu}}\}$. Then, there is a set $\{\ket{\gamma_i^\mu}\}_i\subset\mathcal{H}_\mu$ such that $\{\ket{\uplambda_j^{\mu}}\}_j\bigcup\{\ket{\gamma_i^\mu}\}_i$ comprises a basis for $\mathcal{H}_\mu$. The vector $\ket{\psi}$ can be chosen from $\text{span}\{\ket{\gamma_i^\mu}\}_i$. It is direct to examine that there is some $\ket{\psi}$ in the null space of $\rho_\mu$, i.e., there are $\langle\uplambda_j^\mu\ket{\psi}=0$ for all $j$ as in Eq. (\ref{orth}).
\bibitem{Nuutila} E. Nuutila and E. Soisalon-Soininen, Inf. Process. Lett. \textbf{49}, 9 (1994).
\bibitem{Watrous} J. Watrous, Phys. Rev. Lett. \textbf{93}, 010502 (2004).


\end{thebibliography}
\end{document}